\documentclass[journal=nalefd,manuscript=letter]{achemso}
\usepackage{bm}
\usepackage{multirow}
\usepackage{rotating}
\usepackage{mciteplus}
\usepackage{footmisc}

\author{Liujiang Zhou}
\affiliation{Bremen Center for Computational Materials Science, University of Bremen, Am Falturm 1, 28359 Bremen, Germany}
\email{liujiang86@gmail.com}
\author{Liangzhi Kou}
\affiliation{Integrated Materials Design Centre, School of Chemical Engineering, University of New South Wales, Sydney, New South Wales 2052, Australia}
\author{Yan Sun}
\affiliation{Max Planck Institute for Chemical Physics of Solids, Noethnitzer Str. 40, 01187 Dresden, Germany}
\author{Claudia Felser}
\affiliation{Max Planck Institute for Chemical Physics of Solids, Noethnitzer Str. 40, 01187 Dresden, Germany}
\author{Feiming Hu}
\affiliation{Bremen Center for Computational Materials Science, University of Bremen, Am Falturm 1, 28359 Bremen, Germany}
\author{Guangcun Shan}
\affiliation{Department of Physics and Materials Science and Center for Functional Photonics, City University of Hong Kong, Hong Kong SAR}
\author{Sean C. Smith}
\affiliation{Integrated Materials Design Centre, School of Chemical Engineering, University of New South Wales, Sydney, New South Wales 2052, Australia}
\author{Binghai Yan}
\affiliation{Max Planck Institute for Chemical Physics of Solids, Noethnitzer Str. 40, 01187 Dresden, Germany}
\alsoaffiliation{Max Planck Institute for the Physics of Complex Systems, Noethnitzer Str. 38, 01187 Dresden, Germany}
\email{yan@cpfs.mpg.de}
\author{Thomas Frauenheim}
\affiliation{Bremen Center for Computational Materials Science, University of Bremen, Am Falturm 1, 28359 Bremen, Germany}

\title[An \textsf{achemso} demo]
  {New Family of Quantum Spin Hall Insulators in Two-dimensional Transition-Metal Halide with Large Nontrivial Band Gaps}

\abbreviations{Quantum  spin  Hall  insulator, Two-dimensional Transition-Metal Halide, Gapless edge states, Band inversion, First-principles calculations}
\keywords{Quantum  spin  Hall  insulator, Two-dimensional Transition-Metal Halide, Gapless edge states, Band inversion, First-principles calculations}

\begin{document}

\begin{abstract}
Topological insulators (TIs) are promising for achieving dissipationless transport devices due to the robust gapless states inside the insulating bulk gap. However, currently realized 2D TIs, quantum spin Hall (QSH) insulators, suffer from ultra-high vacuum and extremely low temperature. Thus, seeking for desirable QSH insulators with high feasibility of experimental preparation and large nontrivial gap is of great importance for wide applications in spintronics. Based  on  the  first-principles  calculations, we predict a novel family  of two-dimensional (2D) QSH insulators in transition-metal halide MX (M = Zr, Hf; X = Cl, Br, and I) monolayers, especially, which is the first case based on transition-metal Halide-based QSH insulators. MX family has the large nontrivial gaps of 0.12$-$0.4 eV, comparable with bismuth (111) bilayer (0.2 eV), stanene (0.3 eV) and larger than ZrTe$_5$ (0.1 eV) monolayers and graphene-based sandwiched heterstructures (30$-$70 meV). Their corresponding 3D bulk materials are weak topological insulators from stacking QSH layers, and some of bulk compounds have already been synthesized in experiment. The mechanism for 2D QSH effect in this system originates from a novel \emph{d}$-$\emph{d} band inversion, significantly
different from conventional band inversion between \emph{s}$-$\emph{p}, \emph{p}$-$\emph{p} or \emph{d}$-$\emph{p} orbitals. The realization of pure layered MX monolayers may be prepared by exfoliation from their 3D bulk phases, thus holding great promise for nanoscale device applications and stimulating further efforts on transition metal-based QSH materials.

\end{abstract}

Topological insulators (TIs), characterized with their gapless edge states inside a bulk energy gap, which are topologically protected by time reversal symmetry and from backscattering, show a promising potential application in future dissipationless electronic devices.\cite{hasan_colloquium_2010,qi_topological_2011,yan_topological_2012} Theoretically predicted and experimentally observed TIs usually appear in those materials containing elements with strong spin-orbit coupling, including HgTe/CdTe,\cite{konig_quantum_2007}  InAs/GaSb \cite{knez_evidence_2011} quantum wells, BiSb alloys,\cite{hasan_colloquium_2010} V$_2-$VI$_3$ family compounds (Bi$_2$Se$_3$, Bi$_2$Te$_3$, and Sb$_2$Te$_3$),\cite{zhang_first-principles_2010,zhang_topological_2009} and BiTeI,\cite{bahramy_emergence_2012,wang_magnetotransport_2013} etc. However,  these QSH  insulators showing small bulk gaps  \cite{konig_quantum_2007,knez_evidence_2011} are usually  observed  in experiment only at ultra-high vacuum and extremely low temperature due to weak spin-orbit coupling (SOC). To expand and advance practical application of QSH insulator  or  two dimensional (2D) TIs at room temperature, it is desired to increase the bulk band gap of a TI to overcome the thermal disturbance. Intensive effort has been devoted to enlarge the bulk gaps of  QSH insulators via global structure search (such as Bi$_4$F$_4$\cite{luo_room_2015}), or via creating quantum superlattice,\cite{li_superlattice_2014} or forming 2D III$-$Bi compounds,\cite{chuang_prediction_2014} or via applying external field,\cite{liu_switching_2015} or via structural distortion (1T'-MX$_2$),\cite{qian_quantum_2014} or via substrate effect (such as Bi@Cl-Si(111)\cite{zhou_epitaxial_2014}), or via chemical functionalization (such as stanene,\cite{wu_prediction_2014,li_giant_2015,xu_large-gap_2013} BiX/SbX monolayers,\cite{song_quantum_2014} methyl-munctionalized Bi Bilayer,\cite{ma_robust_2015} etc.); however, since most of them are not layered materials, the well-controlled molecular-beam epitaxy (MBE) technique is usually required to obtain the ultrathin-film samples, giving rise to the difficulty in experimental access. Therefore, desirable materials preferably with large bulk gaps and high feasibility of experimental realization are still extremely scarce and deserve to be explored in experiment and theory.

Nowadays, intensive studies of rare-earth or transition-metal-based QSH insulators,\cite{go_correlation_2012,werner_interaction-driven_2013,weng_topological_2014} showing beyond \emph{s}$-$\emph{p} band inversion \cite{weng_transition-metal_2014} due to the strong electronic interaction instead of SOC,\cite {yang_d-p_2014} greatly enrich the family of QSH insulators and stimulate the further studies on interesting phenomena, such as transition in correlated Dirac fermions \cite{yu_mott_2011} and interaction induced topological Fermi liquid.\cite{castro_topological_2011} The numbers of transition metal atoms-based TIs are extremely rare, and only several examples are reported, that is, square-octahedral lattices of MX$_2$ isomers,\cite{qian_quantum_2014,sun_graphene-like_2015} 2D layered MTe$_5$(M = Zr, Hf) \cite{weng_transition-metal_2014} with the band gaps of 12$-$64 meV and 0.1 eV, respectively. However, these TIs show relatively high formation energies and complicated atomic structures, which may give rise to the difficulty in experimental preparation.

Here, we  report that a serials of \emph{d}$-$\emph{d} band inversion 2D QSH insulators in MX (M=Zr, Hf; X=Cl, Br, I) monolayers, which is distinctive from conventional TIs (such as Bi$_2$Se$_3$ and BiTeI with only \emph{s}$-$\emph{p} band-inversion process). MX monolayer has the simplest stoichiometric ratio and its interlayer binding energy is comparable to other layered systems that have been successfully exfoliated, such as graphite and  MoS$_2$. Hence, the MX monolayer could be obtained via the mechanical exfoliation from the 3D bulk phase as like producing  graphene from graphite. All these MX monolayers are robust QSH insulators against external strain, showing very large tunable nontrivial gaps  in the range of 0.12$-$0.4 eV, comparable with bismuth(111) bilayer (about 0.2 eV) \cite{murakami_quantum_2006}, stanene (0.3 eV) \cite{xu_large-gap_2013} and larger than ZrTe$_5$ (0.1 eV) monolayers \cite{weng_transition-metal_2014} and recent sandwiched graphene-based heterstructures (30$-$70 meV).\cite{kou_graphene-based_2013, kou_robust_2014} ZrCl, ZrI, and HfCl monolayers can also be applied a proper biaxial in-plain stress to further enlarge the nontrivial gaps. Interestingly, a novel band inversion between the pure \emph{d}$-$\emph{d} orbitals is found, greatly enriching the family of TIs. The stability, electronic properties, band inversion, and topological edge states are also discussed.

Layered compouds ZrCl and ZrBr have been synthesized in experiment.\cite{adolphson_crystal_1976,daake_zirconium_1977, marchiando_electronic_1980} Both of them have the closely packed quadruple layers (QLs) each consisting  of tightly bound double hexagonal Zr atomic layers  sandwiched  between two hexagonal halogen atomic layers in the layering sequence of X-M-M-X(Figure 1a). The strong bonding within the slabs is manifest in their high thermal stabilities and melting points above 1100 $^{\circ}$C.\cite{marchiando_electronic_1980} They crystallize in the hexagonal layered structure with space group R$\overline{3}$m. Due to the strong chemical bonding interaction within each four-layer  atomic layer, the adjacent layers are only weakly coupled via the van der Waals' (vdW) interaction. The interlayer binding energy of ZrBr monolayer is 13.1 meV/$\AA^2$, which is as weak as that of graphite (12 meV/$\AA^2$), and is much smaller than that of the Bi$_2$Se$_3$ (27.6 meV/$\AA^2$)\cite{weng_transition-metal_2014} and MoS$_2$ (~26 meV/$\AA^2$).\cite{zhou_large-gap_2014} Just as like the easy procedure of making graphene from graphite simply by exfoliation,\cite{geim_rise_2007,novoselov_electric_2004} the comparably weak interlayer binding energy suggests that a single layer of ZrBr (or other MX) may be formed in a similar flexible and efficient approach.

To study the QSH states of MX monolayers, it is very crucial to judge their corresponding topological state of 3D bulk phases stacked via weak interlayer coupling by three free-standing single layers (Figure 1a). As shown in Figure S1, band dispersion along the $\Gamma$$-$$Z$ direction is almost the same, confirming the feature of weak interlayer coupling. Before applying the SOC, the band structure show it is a semimetal (Figure S1a); while, when apply SOC, the energy gap of 10 meV (PBE result) or 0.22 eV (HSE06) is opened up by the SOC splitting of the degenerated Zr-\textit{d} states (Figure S1b and Table 1). In Figure 1b,  eight time reversal invariant momenta (TRIM) $\Gamma_{i}$ $(i=1, 2,..., 8)$ are labeled as $\Gamma$, $Z$, $M_{1,2,3}$, and $L_{1,2,3}$. According to Fu and Kane¡¯s method,\cite{fu_topological_2007} the parity eigenvalues $\delta$ of valence bands are as follows: $\delta_{\Gamma}$ = $\delta_{Z}$ = $-1$, $\delta_{M_{1,2,3}}$ = $\delta_{L_{1,2,3}}$ = +1. In 3D bulk materials, the main topological index $\delta_i$ of $Z_2$ ($\nu_0; \nu_1\nu_2\nu_3$) is determined by $(-1)^\nu_0$ = $\Pi_{i}\sigma_i$. The other three indices are determined as $(-1)^\nu_1$ = $\sigma_{M_1}\sigma_{M_2}\sigma_{L_1}\sigma_{L_1}$, $(-1)^\nu_2$ = $\sigma_{M_2}\sigma_{M_3}\sigma_{L_2}\sigma_{L_3}$, and $(-1)^\nu_3$ = $\sigma_{L_1}\sigma_{L_2}\sigma_{L_3}\sigma_Z$. Namely, $\nu_0$ corresponds to all eight TRIM; $\nu_{1,2,3}$ correspond to four TRIM on a surface, not inclusive of the $\Gamma$ point, where the surface belongs to the parallelepiped [see Fig. 1(b)] formed by these eight TRIM. Thus, 3D bulk ZrBr belongs to the topological index $Z_2$ (0; 001) class of weak TIs.  This weak TI feature in 3D ZrBr is consistent with the case of layered KHgSb,\cite{yan_prediction_2012} stacked by its corresponding 2D TI layers  with odd-layered structures. This issue on the topological feature of monolayer  structure will be systematically investigated in the following.

Extending ZrCl, ZrBr monolayers to their homogeneous MX(M = Zr, Hf; X = Cl, Br, I) monolayers, they have closely related structures (Figure 1c$-$d) and very similar electronic properties. The optimized lattice constants are listed in Table 1. The lattice constant has an increase trend from 3.45 to 3.70 and  3.40 to 3.65 {\AA} as the increasing of atomic radius of X atom in ZrX and HfX monolayers, respectively. While, when fixed halogen atom, the lattice constant have a reverse variation trend (decrease about 1.5 \%) due to the fact that atomic radius of Zr and Hf are influenced by lanthanide contraction effect.\cite{housecroft_inorganic_2008} (hereafter, we take ZrBr as an example). ZrBr monolayer is a  four-layer  atomic sheets close-packed in the order of Br-Zr-Zr-Br, consisting of double Zr atomic layers  sandwiched  between two  Br atom layers, also showing the  $D3d$(R$\overline{3}$m) symmetry. The calculated phonon spectrums(Figure S2) show  no negative frequency for MX monolayers, which suggests it is a stable phase without any dynamical instability, related to the interactions between atomic layers

The  electronic  band  structures  of  MX monolayers based on parametrized by the Perdew, Burke, and Ernzerhof (PBE) \cite{perdew_self-interaction_1981} exchange correlation interaction are shown in Figure. 2 and Figure S3. In the cases without SOC, all these MX monolayers show semimetal feature. Notably, the two energy bands cross  linearly  at  the  $\Gamma$  point,  suggesting these materials can be considered as a gapless semiconductor, or alternatively, as a semi-metal  with  zero  density  of  states at  Fermi  level for ZrBr and HfBr monolayers and with small amounts DOS for others.  In the vicinity of Fermi level ($E_F$), bands including valence and conduction bands mostly consist of M-\emph{d}$_{xy}$, and M-\emph{d}$_{x^2-y^2}$ orbitals according to the partial band projections. As long as the SOC is taken into consideration, all these degenerated M-\emph{d}$_{xz}$ and M-\emph{d}$_{yz}$ orbitals (denoted as $\emph{d}_I$ orbital) and M-$\emph{d}_{xy}$ and M-\emph{d}$_{x^2-y^2}$ orbitals (denoted as $\emph{d}_{II}$ orbital) are lifted out and splited into two single states, opening the energy gaps to 0.03, 0.10 and 0.03 eV for ZrBr, HfBr and HfI monolayers, respectively.

 Although including SOC will open up the band gaps for ZrCl, HfCl monolayers at the $\Gamma$  point, conduction band minimum (CBM) locates below the $E_F$ and thus leads to the semimetal feature (Figure S3d and f). While for the case of ZrI monolayer, a gap of 20 meV can only be opened up  at the  $\Gamma$ point, and the $E_F$ still pass through the conduction band in the vicinity of $\Gamma$  point, also suggesting a semimetal property (see Figure S3e). We have investigated the $Z_2$ topological invariant \cite{kane_$z_2$_2005,kane_quantum_2005} of MX monolayers by evaluating the parity eigenvalues of occupied states at four time-reversal-invariant-momentum(TRIM) points of the BZ,\cite{fu_topological_2007} and concluded that all these six cases are nontrivial QSH insulators with $Z_2$ = 1. Since the PBE functional is known to usually underestimate the band gap, we have performed additional calculations for MX monolayers using hybrid functional (HSE06) \cite{heyd_erratum,heyd_hybrid_2003} to correct the band gaps (Figure S3g$-$i). The nontrivial gaps of ZrBr, HfBr and HfI monolayers have an enlarged nontrivial gaps by about 0.2 eV. Interestingly, ZrCl, ZrI and HfCl monolayers also show QSH insulator features with the band gaps of 0.12, 0.12 and 0.21 eV, respectively, which is a big difference from the PBE results. The tunable HSE06 band gaps in the range of $0.12-0.40$ eV, are comparable with bismuth (111) bilayer, (about 0.2 eV)\cite{murakami_quantum_2006} stanene (0.3 eV) \cite{xu_large-gap_2013} and larger than ZrTe$_5$ (0.1 eV) monolayers \cite{weng_transition-metal_2014} and recent sandwiched graphene-based heterstructures (30$-$70 meV).\cite{kou_graphene-based_2013, kou_robust_2014} The comparatively large tunable nontrivial gaps in a pure monolayer materials without chemical adsorption, or field effects, which are very beneficial for the future experimental preparation for MX monolayers via simple exfoliation like graphene from its 3D bulk phase and makes them highly adaptable in various application environments.

It should be emphasized that though ZrCl and HfCl monolayers show  QSH semimetal feature from PBE results (Figure S3 and Table 1), it may  be available to push up the CBM and thus to achieve the topological phase transition from QSH semimetal to QSH insulator. Since the main component of CBM at $M$  point is from \emph{d}$_{x^2-y^2}$ and \emph{d}$_{z^2}$ orbitals, the CBM level can be modulated by applying stretching strain within the in-plane x and y direction simultaneously. The larger strain is applied, the longer the M-X bonds length and the weaker the orbital hybridization between $d$ orbitals, finally leading to the CBM levels shift up. To verify this assumption, by applying a $6 \%$ biaxial stretching strain on ZrCl and HfCl monolayers, we can find that the CBMs at M point shift upwards and finally locate above Fermi level for ZrCl and HfCl monolayers (Figure S4). The CB levels near the $\Gamma$ points also are pushed up further, and thus these band gaps are opened up to 0.03, 0.15 eV (PBE results) for ZrCl and HfCl monolayer, respectively, leading to a topological phase transition successfully. The HSE06 calculation will further enlarge their nontrivial band gaps to 0.21 and 0.41 eV, respectively, which are relatively larger than those (0.12 and 0.21 eV) of their equilibrium geometries. This approach is also applied to HfBr monolayer to increase the nontrivial gap due to the similar CBM energy level and orbital components at the $M$  point.

The low energy physics of MX monolayers can be described via a tight-binding (TB) model based on Slater-Koster (SK) approximation.\cite{slater_simplified_1954} From orbital-resolved band structures in Figure 2a, the Bloch states of monolayer ZrBr near the Fermi level mostly consist of Zr-$\emph{d}_{I}$ ($\emph{d}_{xy}$, \emph{d}$_{x^2-y^2}$) orbital and small amounts of $\emph{d}_{II}$ ($\emph{d}_{xz}$, $\emph{d}_{yz}$) orbitals. Moreover, ZrBr monolayer belonging the point group of D3h splits the Zr$-$d orbitals into three categories: $A_1g$($\emph{d}_{z^2}$, $E_g$ ($\emph{d}_{xy}$, \emph{d}$_{x^2-y^2}$) and $E_g$ ($\emph{d}_{xz}$, $\emph{d}_{yz}$), where $A_1g$ and $E_g$ are the Mulliken notations for the irreducible representations (IRs). Thus, to describe the low energy physics, we construct a tight-binding model based on the bases $\{\emph{d}_{xy}$, \emph{d}$_{x^2-y^2}$, $\emph{d}_{xz}$, $\emph{d}_{yz}\}$.
\begin{equation}
 H_{TB}=\sum\limits_{i,\alpha}\epsilon_i^{\alpha}c_i^{\alpha+}c_i^{\alpha} + \sum\limits_{<i,j>,\alpha,\beta}t_{ij}^{\alpha\beta}(c_i^{\alpha+}c_j^{\beta})+h.c.)
\end{equation}

Here, $\epsilon_i^{\alpha}$, $c_i^{\alpha+}$, and $c_i^{\alpha}$ represent the on-site energy, creation, and annihilation operators of an electron at the $\alpha$-orbital of the $i$th atom, respectively. The on-site energies of Zr-d orbitals are set to $\epsilon^{xy}$ = $\epsilon^{x^2-y^2}$ = 1.61 eV, $\epsilon^{xz}$ = $\epsilon^{yz}$ = 2.34 eV, respectively. The $t_{ij}^{\alpha\beta}$ parameter is the hopping energy of an electron between $\alpha$-orbital of $i$th M atom and $\beta$-orbital of $j$th M atom ($\alpha$, $\beta$ $\in$ {$\emph{d}_{xy}$, \emph{d}$_{x^2-y^2}$, $\emph{d}_{xz}$, $\emph{d}_{yz}$}, which are obtained by fitting the DFT data according to TB theory (see in Supporting Information). The TB model reproduces well the bands nearest to the Fermi level, especially the order of $\emph{d}_{xy}$, \emph{d}$_{x^2-y^2}$, $\emph{d}_{xz}$, $\emph{d}_{yz}$, as shown in Figure S5a. We also find that $\emph{d}_{I}$ and $\emph{d}_{II}$ are always degenerated at $\Gamma$ point, respectively, which is agreement the DFT  results and symmetry feature of point group $D3h$. By involving a SOC term in the TB Hamiltonian, the band dispersion in the region near the Fermi level can be also well reproduced (Figure S5b). The degenerated d orbital levels are lifted out by $\Delta_{SOC}$ = 2$\lambda$ ($\lambda$ is the SOC strengths of Zr atom) in the opposite direction at $\Gamma$ point, leading to a SOC band gap of 0.03 eV is obtained with $\lambda$ = 0.015 eV.

To illustrate the band inversion process explicitly, we start from Zr-\emph{d} atomic orbitals and consider the effect of chemical bonding and SOC on the energy levels at the $\Gamma$ point for ZrBr monolayer.  This is schematically illustrated in two stages (I), (II) and (III) in Figure 3. Stage (I) represents the chemical bonding process between Zr atoms, during which states around the Fermi energy are mainly contributed by degenerate $\emph{d}_{I}$ orbitals with parity $p$ = +1 and $\emph{d}_{II}$ orbitals with parity $p = -1$, which is indicated by these orbital energy levels and parities of this model with a stretching lattice constant of $170\%$ $a$. At this moment, $\emph{d}_{I}$ orbitals are located below Fermi levels due to the only two \textrm{d} electrons participating the bonding process. In stage II, two Zr atoms move further closely to each other, being equivalent to compressing further the lattice, leading to the degenerate bonding \emph{d}$_{I}$ orbitals shifts upward with respect to the degenerate bonding \emph{d}$_{II}$ orbitals. Because of the strong Zr$-$Zr atomic interaction, band inversion occurs between the $\emph{d}_{I}$ orbitals with parity $p = +1$ and  $\emph{d}_{II}$ orbitals with parity $p = -1$ at a critical stretching lattice of about $140\%$ $a$, inducing a topological phase transition from a TI to a trivial insulator. The band crossings occurring between the $\emph{d}_{z^2}$ orbitals and other $\emph{d}$ orbitals do not contribute to the band inversion because the parity of bonding $\emph{d}_{z^2}$ orbital is always positive  whether it is occupied or not. It is noteworthy that the band inversion is purely from \emph{d}$-$\emph{d} orbitals, different from conventional band inversion from \emph{s}$-$\emph{p} orbitals, or \emph{p}$-$\emph{p} orbitals. In stage III, when including the effect of SOC on equilibrium structure,  the degenerate orbitals are lifted out and their parities remain unchanged, which is also confirmed by the same $Z_2$ invariant of ZrBr monolayer with or without SOC (Table 1). Thus, the band inversion does not originate from the SOC. The role of SOC is only to open up the band gap, which is similar to graphene, ZrTe$_5$ and square-octagonal MX$_2$ structures.\cite{sun_graphene-like_2015,ma_quantum_2015}

The 2D nontrivial insulating states in MX monolayers should support an odd number of topologically protected gapless conducting edge states connecting the valence and conduction bands of each system at certain $k$-points. To see these topological features explicitly, we perform calculations of the edge states by cutting 2D monolayer into nanoribbon with the Wannier functions \cite{marzari_maximally_1997,mostofi_wannier90:_2008} extracted from $ab$ $initio$ calculations(see Methods). In order to eliminate the coupling between two edges, the width of the slabs are up to 70 unit cells for cases. The calculated results of three QSH insulator based on PBE plus SOC presented in Figure 4 show that a pair of gapless edge states are present inside the 2D QW gap at both left and right edges. The Dirac cones are all located at the $\Gamma$ point for three cases.  Although the details depend on different compounds and edges, the non-trivial $Z_2$ invariant guarantees the edge bands always cutting Fermi level odd times. At a given edge, two counter-propagating edge states display opposite spin-polarizations, a typical feature of the 1D helical state of a QSH phase.

To facilitate the experimental preparation of MX monolayers, it is very crucial to find a suitable substrate to hold the MX monolayers via weak interlayer interaction, hoping the substrate have little effect on the electronic structure (including the band gap and topological order) of MX monolayers. The nontrivial band topology is from pure \emph{d}$-$\emph{d} band inversion in double M atomic layers within the QLs, so that the negative substrate effect can be avoided in MX QSH insulators. Take ZrBr monolayer as an example, we find that it can match well with the clean MoTe$_2$(001) (Te-terminated) surface with a very small lattice mismatch ($\sim$0.8 \%). The distance between ZrBr monolayer and MoTe$_2$(001) surface is 3.5 \AA, indicating the weak van der Waals interlayer coupling. The states of substrates are located far from key region near the Fermi-level consisting of various d-orbitals, and the orbital hybridizations between substrate and MX monolayer are almost negligible and the nontrivial band topological properties are still maintained (see in Figure S6). So, it is feasible to observe the nontrivial topological phase in the MX monolayers on a substrate in experiment.

In summary,  MX (M = Zr, Hf; X = Cl, Br, and I) monolayers constitute a novel family of robust QSH insulators, showing  very large tunable nontrivial gaps  in the range of 0.12$-$0.4 eV, which are comparable with bismuth (111) bilayer (0.2 eV), stanene (0.3 eV) and larger than ZrTe$_5$ (0.1 eV) monolayers and graphene-based sandwiched heterstructures (30$-$70 meV). MX family is also the first case based on transition-metal Halide-based QSH insulators. By applying a proper in-plain stress on ZrCl, ZrCl, and HfBr monolayers, it is available to enlarge the nontrivial gaps by shifting upwardly the CBM. The mechanism for the QSH effect is from the band inversion between metal atomic \emph{d}$_{I}$ (degenerated \emph{d}$_{yz}$, \emph{d}$_{xz}$) orbitals and \emph{d}$_{II}$ (degenerated \emph{d}$_{xy}$, \emph{d}$_{{x}^2-{y}^2}$) orbitals, a typical \emph{d}$-$\emph{d} orbitals band inversion, which is different from conventional band inversion from \emph{s}$-$\emph{p}, or  \emph{p}$-$\emph{p} or \emph{d}$-$\emph{p} orbitals. The role of SOC is only to open up the band gap just like graphene. These comparable large nontrivial gaps in pure monolayer materials without chemical adsorption, or applying strain, or distortion, which are very beneficial for the future experimental preparation for MX monolayers via simple exfoliation from its 3D bulk phase, makes them highly adaptable in various application environments. These interesting results may
stimulate further efforts on 2D transition-metal-based QSH insulators.

\section{Methods}

First-principles calculations based on the density functional theory (DFT) are carried out using the Vienna $ab$ $initio $ Simulation Package (VASP) \cite{kresse_efficient_1996}. The exchange correlation interaction is treated within the generalized gradient approximation(GGA) \cite{perdew_generalized_1996}, which is parametrized by the Perdew, Burke, and Ernzerhof (PBE)\cite{perdew_self-interaction_1981}. All the atoms in the unit cell are fully relaxed until the force on each atom is less than 0.01 eV/{\AA}. The interlayer binding energy is calculated according to its definition as the unit-area total-energy difference between the single-layer sheet and the 3D bulk.\cite{lind_structure_2005} Since DFT methods often underestimate the band gap, the screened exchange hybrid density functional by Heyd-Scuseria-Ernzerhof (HSE06) \cite{heyd_erratum,heyd_hybrid_2003} is adopted to correct the PBE band gaps. The tight binding matrix elements are calculated by projection
Bloch states onto maximally localized Wannier functions (MLWFs) \cite{marzari_maximally_1997,mostofi_wannier90:_2008}, using the VASP2WANNIER90 interface \cite{franchini_maximally_2012}. In the MX monolayer, since the bands around Fermi level are almost consisted by these five M-\emph{d} orbitals, the maximally localized Wannier functions (MLWFs) are derived from atomic \emph{d} orbitals. The tight binding parameters are determined from the MLWFs overlap matrix. The phonon calculations are carried out by using the density functional perturbation theory (DFPT)\cite{baroni_phonons_2001} as implemented in the PHONOPY code.\cite{baroni_phonons_2001} The topological character is calculated according to the procedure outlined in Ref. \cite{fu_topological_2007} The central quantity here is the parity product ${\delta_i=\prod_{m=1}^{N}\xi_{2m}(\Gamma_i)}$, where $N$ is the number of the occupied bands and $\xi_{2m}(\Gamma_i)$ is the parity eigenvalue of the 2$m$-th occupied band at the time reversal point $\Gamma_i$. The $Z_2$ invariant ($\nu_0=0$ or 1) is then derived from the product $(-1)^{\nu_0}=\prod_{i}\delta_i$. $\delta_i$ is $-1$ if an odd number of band inversions occur at $\Gamma_i$; otherwise, $\delta_i$ is $+1$. (When the crystal structure has inversion symmetry, $\delta_i$ is the product of parity eigenvalues of the valence bands at $\Gamma_i$.)

\clearpage
\newpage
\begin{suppinfo}
The details of TB model without SOC, TB model with SOC, phonon spectrums, band structures for ZrCl, ZrI and HfCl monolayers, band structures
of ZrCl, HfCl monolayers under $6 \%$ in-plane strain, structure and orbital projection band structure of 2D ZrBr monolayer on the clean MoTe$_2$(001) (Te-terminated) surface. This material is available free of charge via the Internet at http://pubs.acs.org
\end{suppinfo}

\section{Notes}
The authors declare no competing financial interests.

\begin{acknowledgement}
L.Z acknowledges financial support from Bremen University. B.Y. and C.F. acknowledge financial support from the European
Research Council Advanced Grant (ERC 291472). The support of the Supercomputer Center of Northern Germany (HLRN Grant No. hbp00027).
\end{acknowledgement}

\begin{figure*}
\centering
\includegraphics[width=12cm]{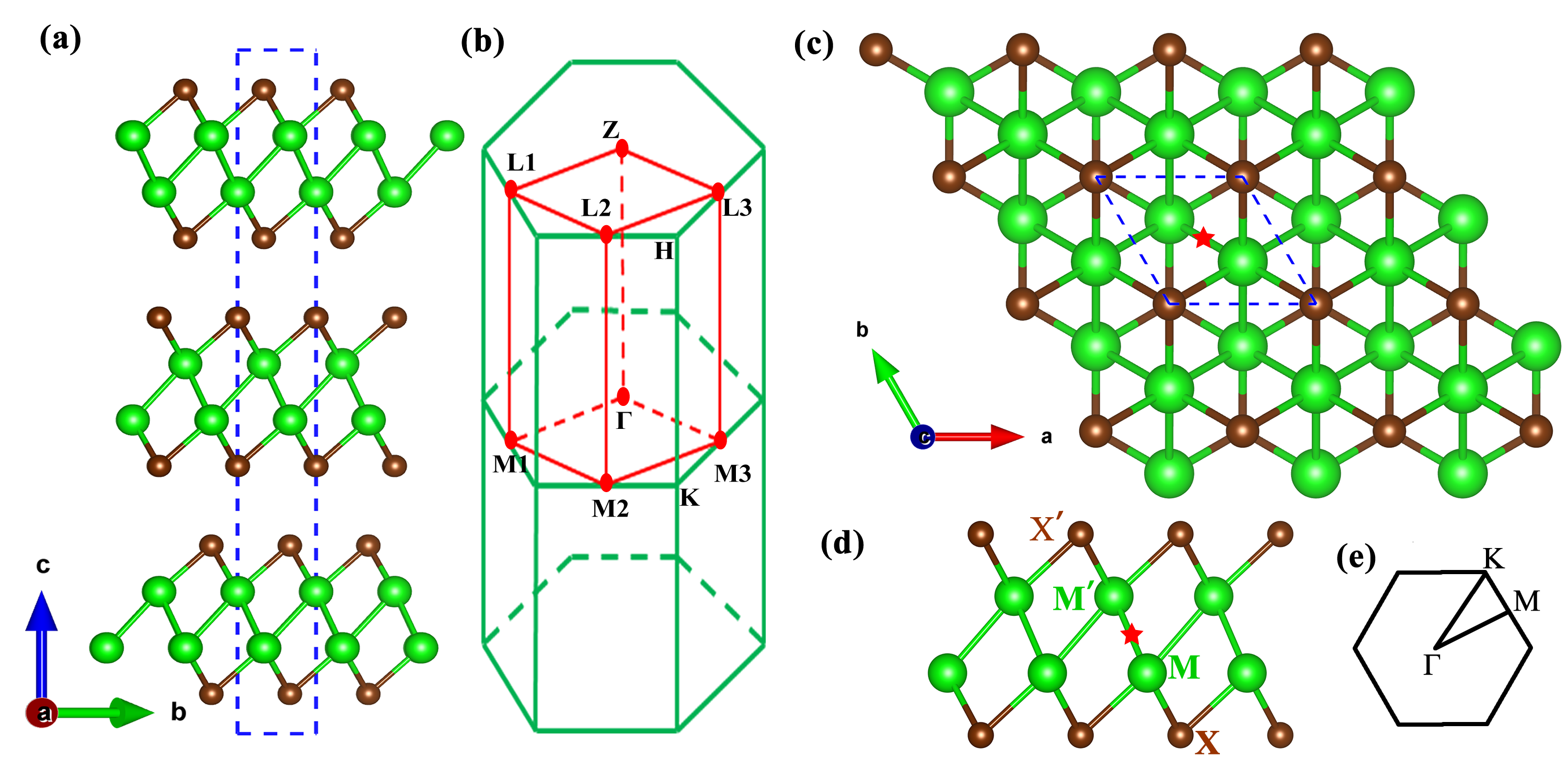}\\
\caption{ (a) The crystal structure of 3D ZrBr. (b) The Brillouin zone (BZ) of  of the honeycomb
lattice in 3D ZrBr crystal. Panels (c) and (d): The top view and side  view of ZrBr monolayer, respectively. In a unit  cell, MX is  related  to M$'$X$'$ by an inversion operation. The inversion center is indicted by Red star in (c) and (d). Panel (e) First Brillouin zone of MX monolayer  and the points of high symmetry.}
\label{Fig. 1}
\end{figure*}

\newpage
\begin{figure}
\centering
\includegraphics[width=14cm]{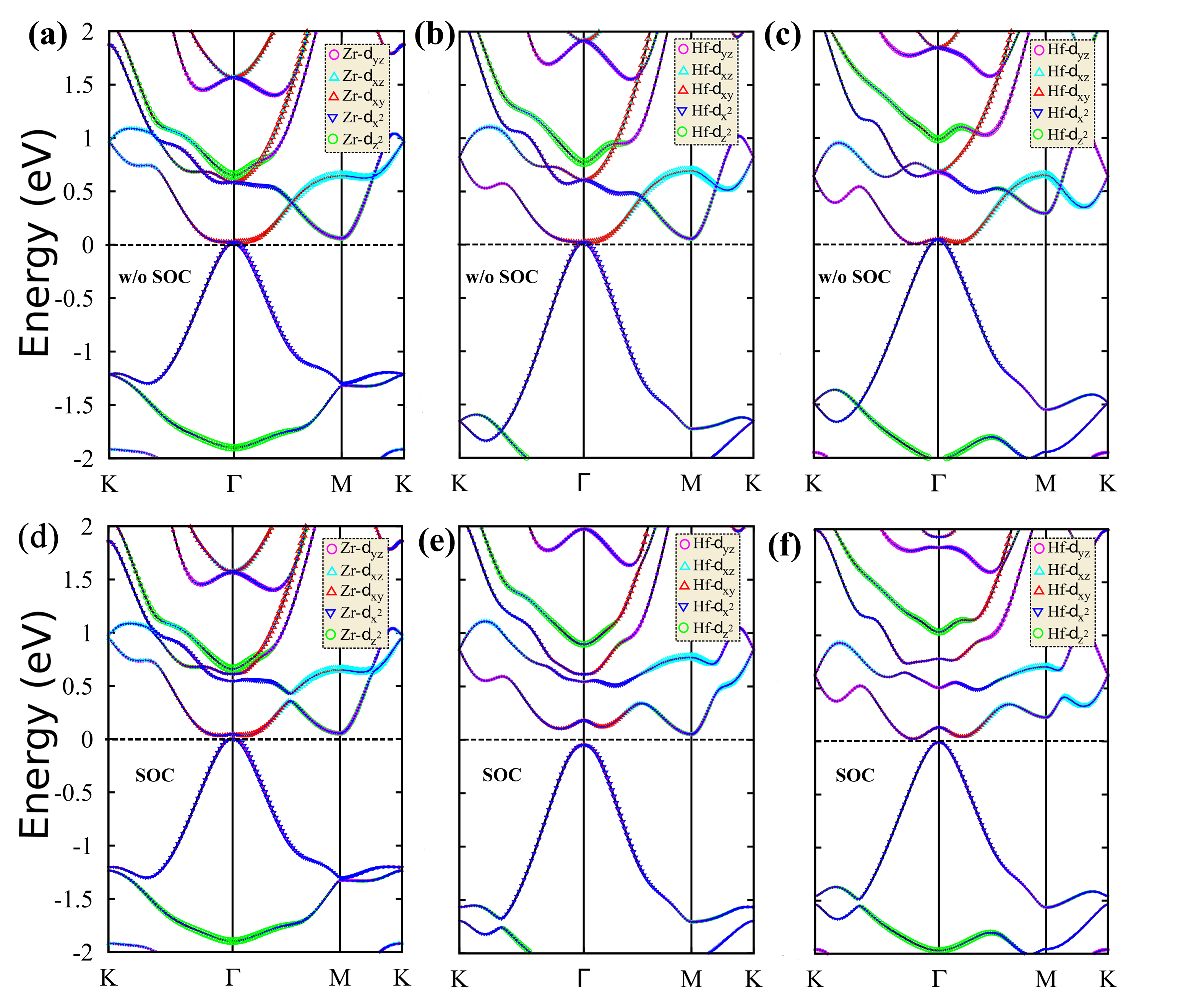}
\caption{The calculated PBE band structures with and without SOC, respectively. Panels (a) and (d): ZrBr  monolayer  without and with SOC, respectively; panels (b) and (e): HfBr monolayer  without and with SOC, respectively; Panels (c) and (f): HfI  monolayer  without and with SOC, respectively. The upper three panels of bands near Fermi energy are mainly dominated M-\emph{d}$_{xy}$ and M-\emph{d}$_{x^2-y^2}$, where M = Zr and Hf. In (e), the energy level of CBM with the dominated orbitals of \emph{d}$_{x^2-y^2}$ at M  point is lower than conduction band level around the $\Gamma$ near Fermi level,  indicting it is available to shift upwards  the CBM by applying in-plane biaxial strain.}
\label{Fig. 2}
\end{figure}

\newpage

\begin{figure}
\centering
\includegraphics [width=10cm]{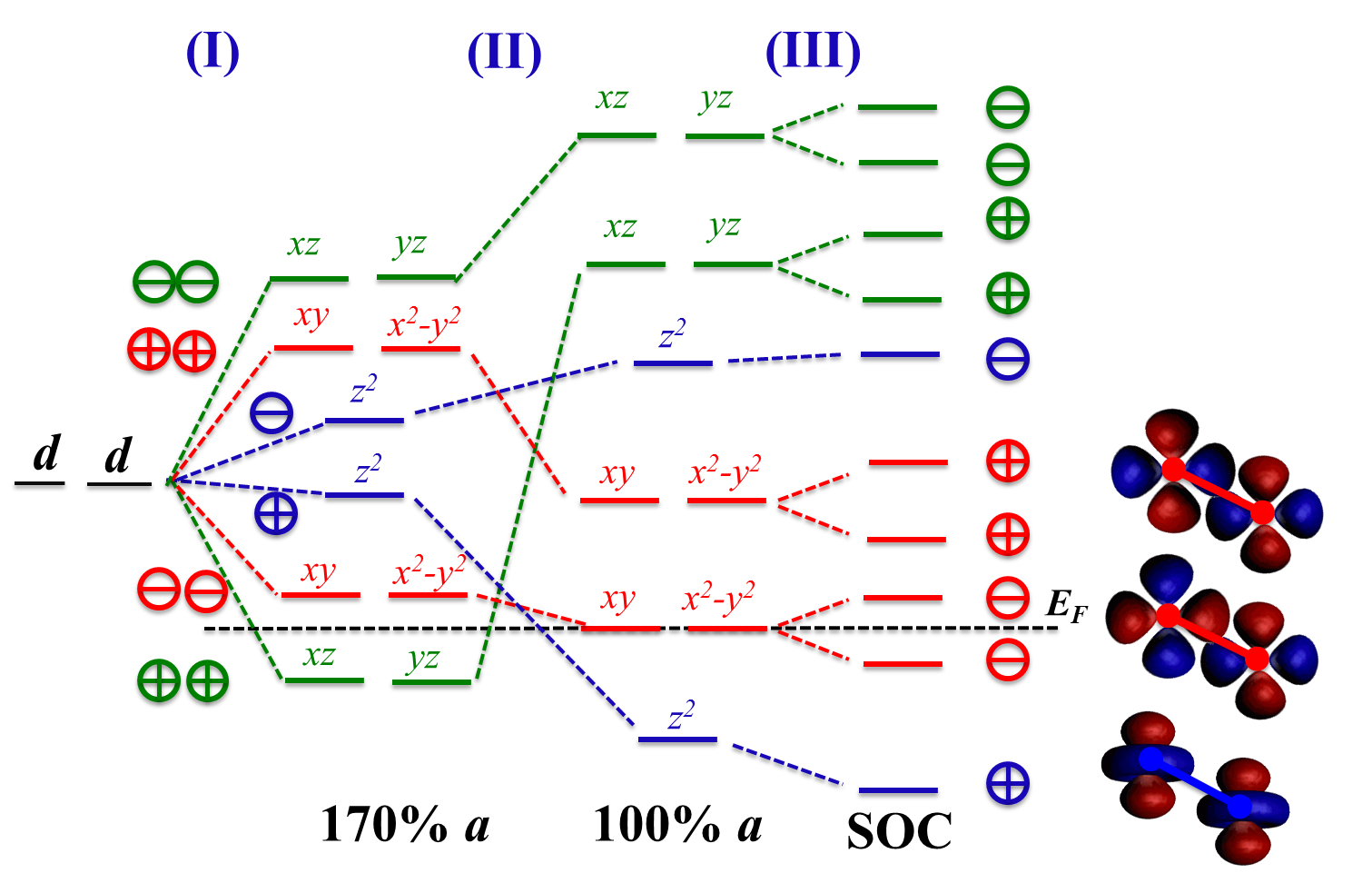}\\
\caption{Schematic illustration of the evolution from the atomic orbitals  \emph{d}$_{xy}$, \emph{d}$_{yz}$, \emph{d}$_{xz}$, \emph{d}$_{{x}^2-{y}^2}$ and \emph{d}$_{z^2}$ of Zr into the conduction and valence bands of ZrBr monolayer at the  $\Gamma$ point. The three different stages (I), (II) and (III) represent the effect of chemical bonding, crystal-field splitting and SOC, respectively. The black dashed line represents the Fermi energy.  Parity values are presented near the splitted \emph{d} orbitals. The \emph{d}$_{x^2-y^2}$ and \emph{d}$_{z^2}$ orbital plots at the rightmost are used to display the variation of parity. The band inversion will occur between degenerate \emph{d}$_{xz}$, \emph{d}$_{yz}$ orbitals (denoted as \emph{d}$_{I}$) and \emph{d}$_{xy}$, \emph{d}$_{x^2-y^2}$ (denoted as \emph{d}$_{II}$) orbitals in the process of stretching lattice parameter \emph{a} with the lattice symmetry preserved.}
\label{Fig. 3}
\end{figure}
\newpage

\begin{figure}
\centering
\includegraphics[width=12cm]{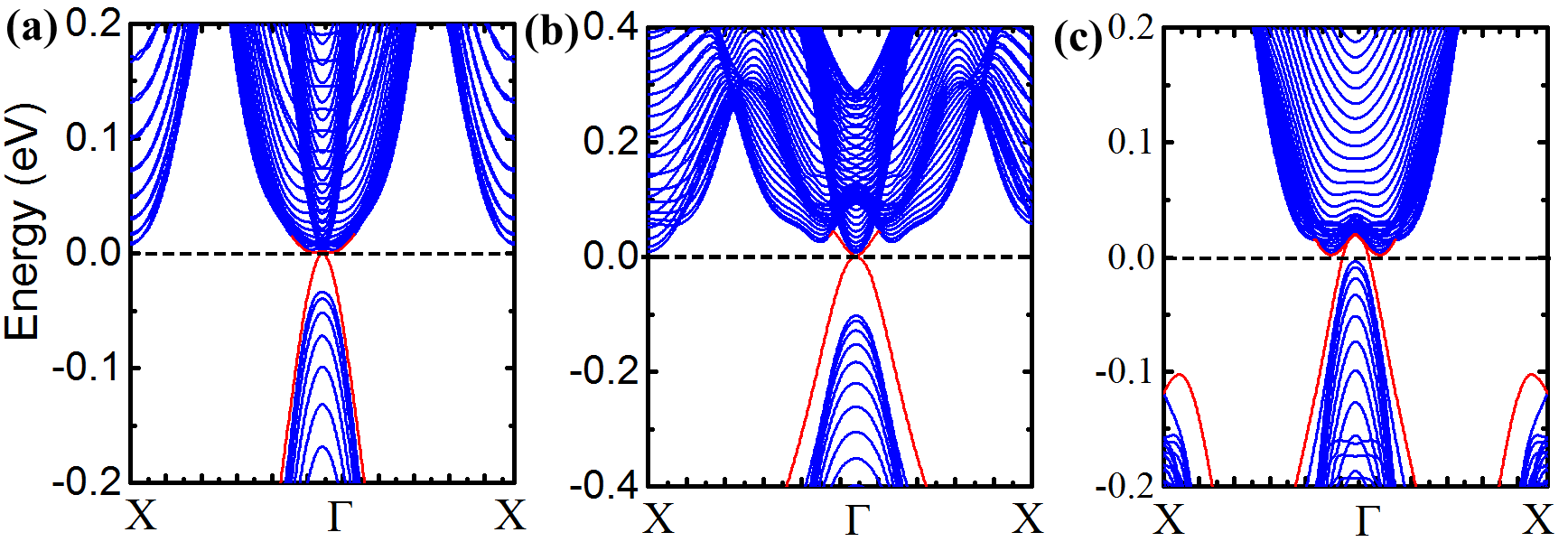}\\
\caption{The calculated topological edge states of  (a) ZrBr, (b) HfBr and (c) HfI monolayers with SOC. The Dirac helical states are denoted by the red solid lines, which exist at the edges of the ribbon structure. The Fermi energy is set to 0 eV.}
\label{Fig. 4}
\end{figure}

\newpage

\begin{table}
\caption{\label{tab:1} The predicted lattice constants of MX monolayers and 3D ZrBr, and their band gaps (eV) with (w) and without (w/o) SOC based on PBE and HSE06.}
\begin{center}
\begin{tabular}{l c c c c c c c}
\hline
 Compound &  ZrCl & ZrBr & ZrI& HfCl & HfBr & HfI & 3D ZrBr \\
\hline
   Lattice a ({\AA})            & 3.45	           & 3.53	 & 3.70	    & 3.38	 &3.48  &	3.65  & 3.50  \\
   PBE gap w/o SOC              & s.m.$^a$         & s.m.    & s.m.     & s.m.   &s.m.  &   s.m.  & s.m. \\
   PBE gap w SOC                & s.m.             & 0.03    &	s.m.    &s.m.	 & 0.10 & 	0.03  & 0.007\\
   HSE$06$ gap w SOC            &	0.12           & 0.21    &.12       & 0.21 	 & 0.40 &	0.29  & 0.22 \\
   $Z_2$  invariant ${\nu_0}$   & 1                & 1       & 1        & 1      &1     &   1     & (0;001)\\
\hline
$^a$ semimetal                  &                   &         &         &         &      &          & \\
\end{tabular}
\end{center}
\end{table}

\bibliography{M-X}

\begin{tocentry}
\includegraphics [width=8.5cm] {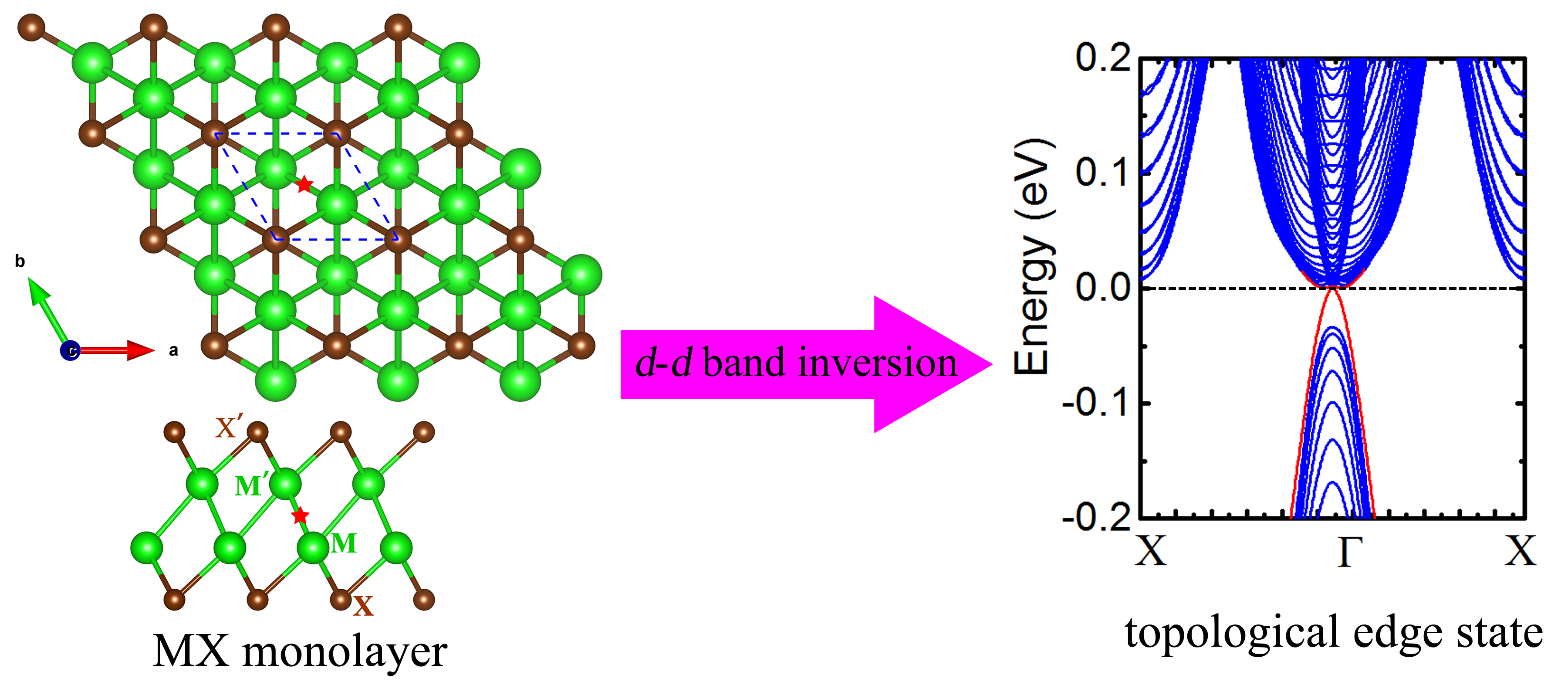}

 Based  on  the  first-principles  calculations, we predict a novel family  of two-dimensional (2D) QSH materials  in Transition-Metal Halide $MX$ (M = Zr, Hf; X=Cl, Br, and I) monolayers with large nontrivial gaps of $0.12-0.4$ eV. A novel $d$-$d$ band inversion is responsible for the 2D QSH effect, distinctive from conventional band inversion between $s$-$p$ orbitals, or $p$-$p$ orbitals.
\end{tocentry}

\end{document}